\documentclass[conference]{IEEEtran}
\usepackage{amsmath,amssymb,bm,cite,algorithm,algorithmic,amsthm}
\usepackage{graphicx, xcolor}

\begin{document}
\title{Low Complexity Channel Estimation for Millimeter Wave Systems with Hybrid A/D Antenna Processing}
\author{
\IEEEauthorblockN{George~C.~Alexandropoulos and Symeon Chouvardas}
\IEEEauthorblockA{Mathematical and Algorithmic Sciences Lab, Huawei Technologies France, 92100 Boulogne-Billancourt, France}
emails: \{george.alexandropoulos, symeon.chouvardas\}@huawei.com}

\maketitle

\begin{abstract}
The availability of large bandwidth at millimeter wave (mmWave) frequencies is one of the major factors that rendered very high frequencies a promising candidate enabler for fifth generation (5G) mobile communication networks. To confront with the intrinsic characteristics of signal propagation at frequencies of tens of GHz and being able to achieve data rates of the order of gigabits per second, mmWave systems are expected to employ large antenna arrays that implement highly directional beamforming. In this paper, we consider mmWave wireless systems comprising of nodes equipped with large antenna arrays and being capable of performing hybrid analog and digital (A/D) processing. Intending at realizing channel-aware transmit and receive beamforming, we focus on designing low complexity compressed sensing channel estimation. In particular, by adopting a temporally correlated mmWave channel model, we present two compressed sensing algorithms that exploit the temporal correlation to reduce the complexity of sparse channel estimation, one being greedy and the other one being iterative. Our representative performance evaluation results offer useful insights on the interplay among some system and operation parameters, and the accuracy of channel estimation.   
\end{abstract}


\section{Introduction}\label{sec:Intro}
Several publications have appeared and numerous panel discussions have taken place on the requirements and candidate technologies for next generation (5G) mobile communication networks \cite{Boccardi_COMmag, Andrews_JSAC}. Although a consensus on the challenges for these networks (including 1 ms end-to-end latency, 10 Gbps throughput, and 1000K network connections per square meter) has been reached, both academia and industry are still working on network architectures and techniques meeting them either one-by-one or in groups of few. Mobile communication in high frequency bands, namely centimeter and millimeter wave (mmWave), is a promising technology for addressing some of the 5G requirements \cite{Rappaport_Access}. Short-range mmWave communication at the unlicensed band of 60 GHz is already standardized in IEEE 802.11ad \cite{J:IEEE_11ad} offering up to 7 Gbps data rate, and initial investigations on mmWave cellular systems have identified their potentials together with their key challenges \cite{Rappaport_Access, Alkhateeb_JSTSP}. 

High frequency mobile communication will rely on directional transmissions and receptions. By exploiting the fact that the wavelength at mmWave frequencies is very small, very large antenna arrays can be packed into small form factors and, thus, can be used to implement highly directional beamforming to support long outdoor links. To achieve the full benefit from beamforming in a multi-antenna communication link, the entire channel state information needs to be available at both communication ends. However, this information is hard to acquire in mmWave systems due to the low coherence time, the radio frequency (RF) hardware limitations \cite{Alkhateeb_JSTSP} (RF chains are usually much less than the number of antenna elements), and the small signal-to-noise ratio (SNR) before beamforming. The majority of the available approaches for mmWave channel estimation and tracking utilizes sets of predefined beams, also known as beam codebooks, at both communication ends. The number of beams that can be realized from a node per time instant depends on the number of its available RF chains. One family of approaches (e$.$g$.$, \cite{Wang_JSAC, J:Gigabit_2013, Hur_TCOM, Hosoya_TAP2015}) is based on beam switching in order to find the pair or pairs of beams between a transmitting and a receiving node resulting in meeting a predefined threshold for data communication. The exhaustive search beamforming protocol for medium access control adopted in the IEEE 802.11ad standard is mainly based on the multi-sector beamforming technique of \cite{Hosoya_TAP2015}. Recently, another family of approaches (e$.$g$.$, \cite{Alkhateeb_JSTSP, Eliasi_ArXiv2014, Marzi_TSP2016, Bazzi_SPAWC2016}) capitalized on the spatial sparsity of mmWave channels \cite{Sayeed_JSTSP, Zhang_WCNC2010} to estimate the entire channel gain matrix or its second-order statistics using tools from the compressed sensing (CS) theory \cite{cande2008introduction}.  

In this paper, we adopt the CS-based formulation of \cite{Alkhateeb_JSTSP} for the mmWave channel estimation problem and devise low complexity algorithms for systems consisting of nodes equipped with large antenna arrays, and being capable of performing hybrid analog and digital (A/D) processing. We consider the temporally correlated mmWave channel model of \cite{HeTaHaKu14} and exploit the temporal correlation to reduce the complexity and duration of sparse channel estimation, without necessarily degrading its performance. In doing so, we assume that the angles of departure and arrival of the signal propagation paths in between successive coherent channel blocks fall within predictable limits, and present one greedy and one iterative CS channel estimation algorithms. One possible way to predict the angle limits is through node position estimation techniques \cite{Taranto_SPmag2014}. We present performance evaluation results for the mean squared error (MSE) of the proposed CS channel estimation and the mean achievable rate with channel-aware beamforming for various system and operation parameters.

\textit{Notation:} Vectors and matrices are denoted by boldface lowercase letters and boldface capital letters, respectively. The conjugate, transpose, and Hermitian transpose of a matrix $\mathbf{A}$ are denoted by $\mathbf{A}^*$, $\mathbf{A}^{\rm T}$, and $\mathbf{A}^{\rm H}$, respectively, while $\mathbf{I}_{n}$ ($n\geq2$) is the $n\times n$ identity matrix. $\|\mathbf{a}\|$ and $\|\mathbf{a}\|_0$ stand for the Euclidean and the $\ell_0$ norm of $\mathbf{a}$, and ${\rm diag}\{\mathbf{a}\}$ denotes a square diagonal matrix with $\mathbf{a}$'s elements in its main diagonal. $[\mathbf{A}]_{i,j}$ and $\mathbf{A}^{(i)}$ represent $\mathbf{A}$'s $(i,j)$th element and $i$th column, respectively. $\mathbf{A}\circ\mathbf{B}$ represents the Khatri-Rao product of $\mathbf{A}$ and $\mathbf{B}$, while $\mathbf{A}\otimes\mathbf{B}$ denotes their Kronecker product. $\mathcal{R}$ and $\mathcal{C}$ represent the real and complex number sets, respectively, whereas $\mathbb{E}\{\cdot\}$ is the expectation operator and ${\rm var}\{\cdot\}$ is the variance operator. Notation $x\sim\mathcal{C}\mathcal{N}\left(0,\sigma^{2}\right)$ indicates that $x$ is a circularly-symmetric complex Gaussian random variable with zero mean and variance $\sigma^{2}$. 

\section{System and Channel Models}\label{sec:System_Model}
We present in the following the system model under investigation as well as the considered wireless channel model suitable for mmWave communication with multi-antenna transceivers.

\subsection{System Model}
Suppose a wireless communication system operating in the mmWave frequency band consisting of a $N_{\rm T}$-antenna transmitter (TX) wishing to send $d$ independent data streams to a $N_{\rm R}$-antenna receiver (RX), where $d\leq\min(N_{\rm T},N_{\rm R})$. Both TX and RX are assumed to be equipped with $N_{\rm RF}$ radio frequency (RF) chains, where $N_{\rm RF}\leq N_{\rm T}$ and $N_{\rm RF}\leq N_{\rm R}$. It must hold $d\leq N_{\rm RF}$ for possible correct decoding at RX. Before transmission, TX linearly processes the symbol vector $\mathbf{s}\in\mathcal{C}^{d\times1}$ with a precoding matrix $\mathbf{V}\in\mathcal{C}^{N_{\rm T}\times d}$. In this work, we assume that $\mathbf{V}$ is decomposed as $\mathbf{V}\triangleq\mathbf{V}_{\rm RF}\mathbf{V}_{\rm BB}$, where $\mathbf{V}_{\rm RF}\in\mathcal{C}^{N_{\rm T}\times N_{\rm RF}}$ denotes the RF precoding matrix implemented using only analog phase shifters (its entries are of constant amplitude) and $\mathbf{V}_{\rm BB}\in\mathcal{C}^{N_{\rm RF}\times d}$ is the baseband (BB) precoding matrix. In our system model, we normalize the entries of $\mathbf{V}_{\rm RF}$ as $|[\mathbf{V}_{\rm RF}]_{i,j}|^2\triangleq N_{\rm T}^{-1}$ with $i=1,2,\ldots,N_{\rm T}$ and $j=1,2,\ldots,N_{\rm RF}$. In addition, we assume that for $\mathbf{V}$ it holds $\|\mathbf{V}^{(n)}\|\triangleq1$ $\forall n=1,2,\ldots,d$. A total power constraint ${\rm P}$ is also considered for TX such that $\mathbb{E}\{\|\mathbf{V}\mathbf{P}^{\frac{1}{2}}\mathbf{s}\|^2\}\leq {\rm P}$, where $\mathbf{P}={\rm diag}\{[P_1\,P_2\,\ldots\,P_d]\}\in\mathcal{R}_+^{d\times d}$ with $P_n$ denoting the power allocated from TX to its $n$th data stream.

The BB $N_{\rm R}\times 1$ complex-valued received signal at RX can be mathematically expressed as
\begin{equation}\label{Eq:System_Model}
\mathbf{r} = \mathbf{H}\mathbf{V}\mathbf{P}^{\frac{1}{2}}\mathbf{s} + \mathbf{n},
\end{equation}
where $\mathbf{H}\in\mathcal{C}^{N_{\rm R}\times N_{\rm T}}$ denotes the channel gain matrix between RX and TX, and $\mathbf{n}\in\mathcal{C}^{N_{\rm R}\times 1}$ represents the zero-mean additive white Gaussian noise (AWGN) vector with covariance matrix $\sigma^2\mathbf{I}_{N_{\rm R}}$. After signal reception, RX is assumed to process $\mathbf{y}$ with a linear filter $\mathbf{U}\in\mathcal{C}^{N_{\rm R}\times d}$ in order to obtain an estimate of the transmitted symbol vector $\mathbf{s}$ as
\begin{equation}\label{Eq:Estimated_Symbols}
\hat{\mathbf{s}} \triangleq \mathbf{U}^{\rm H}\mathbf{r} = \mathbf{U}^{\rm H}\mathbf{H}\mathbf{V}\mathbf{P}^{\frac{1}{2}}\mathbf{s} + \mathbf{U}^{\rm H}\mathbf{n}.
\end{equation}
At RX, $\mathbf{U}$ is assumed to be designed as $\mathbf{U}\triangleq\mathbf{U}_{\rm RF}\mathbf{U}_{\rm BB}$, where $\mathbf{U}_{\rm RF}\in\mathcal{C}^{N_{\rm R}\times N_{\rm RF}}$ denotes the RF filter implemented similar to $\mathbf{V}_{\rm RF}$ (i$.$e$.$, its entries are of constant amplitude)  and $\mathbf{U}_{\rm BB}\in\mathcal{C}^{N_{\rm RF}\times d}$ represents the BB filter. In addition, we normalize the entries of $\mathbf{U}_{\rm RF}$ as $|[\mathbf{U}_{\rm RF}]_{i,j}|^2\triangleq N_{\rm R}^{-1}$ with $i=1,2,\ldots,N_{\rm R}$ and $j=1,2,\ldots,N_{\rm RF}$.

\subsection{Channel Model}\label{Sec:Channel_Model}
Similar to \cite{Sayeed_JSTSP, Zhang_WCNC2010, Alkhateeb_JSTSP, HeTaHaKu14}, we adopt a geometric channel model with $L$ scatterers, where each scatterer contributes a single propagation path in the TX and RX communication link of physical distance $D$. Under this channel model, $\mathbf{H}$ can be expressed as
\begin{equation}\label{Eq:Channel_Model}
\mathbf{H} = \mathbf{A}_{\rm R}\left(\boldsymbol{\theta}\right){\rm diag}\{\boldsymbol{a}\}\mathbf{A}_{\rm T}^{\rm H}\left(\boldsymbol{\phi}\right),
\end{equation}
where matrices $\mathbf{A}_{\rm T}\left(\boldsymbol{\phi}\right)\in\mathcal{C}^{N_{\rm T}\times L}$, with $\boldsymbol{\phi}\triangleq[\phi_1\,\phi_2\,\cdots\,\phi_L]$, and $\mathbf{A}_{\rm R}\left(\boldsymbol{\theta}\right)\in\mathcal{C}^{N_{\rm R}\times L}$, with $\boldsymbol{\theta}\triangleq[\theta_1\,\theta_2\,\cdots\,\theta_L]$, are defined as
\begin{subequations}
\begin{equation}\label{Eq:A_TX}
\mathbf{A}_{\rm T}\left(\boldsymbol{\phi}\right) \triangleq \left[\mathbf{a}_{\rm T}\left(\phi_1\right)\,\mathbf{a}_{\rm T}\left(\phi_2\right)\,\cdots\,\mathbf{a}_{\rm T}\left(\phi_L\right)\right],
\end{equation}
\begin{equation}\label{Eq:A_RX}
\mathbf{A}_{\rm R}\left(\boldsymbol{\theta}\right) \triangleq \left[\mathbf{a}_{\rm R}\left(\theta_1\right)\,\mathbf{a}_{\rm R}\left(\theta_2\right)\,\cdots\,\mathbf{a}_{\rm R}\left(\theta_L\right)\right].
\end{equation}
\end{subequations}
In \eqref{Eq:A_TX} and \eqref{Eq:A_RX}, variable $\phi_\ell\in[0,2\pi]$ with $\ell=1,2,\ldots,L$ denotes the $\ell$th path's angle of departure (AoD) from TX and variable $\theta_\ell\in[0,2\pi]$ represents the $\ell$th path's angle of arrival (AoA) at RX. In addition, $\mathbf{a}_{\rm T}\left(\phi_\ell\right)\in\mathcal{C}^{N_{\rm T}\times 1}$ and $\mathbf{a}_{\rm R}\left(\theta_\ell\right)\in\mathcal{C}^{N_{\rm R}\times 1}$ are the array response vectors at TX and RX, respectively (for uniform linear antenna arrays (ULAs), these vectors are given by \cite[eq. (5)]{Alkhateeb_JSTSP}). In \eqref{Eq:Channel_Model}, $\boldsymbol{a}\in\mathcal{C}^{L\times 1}$ includes the path channel gains $\alpha_\ell$ $\forall\ell=1,2,\ldots,L$. We further assume that each path's amplitude is Rayleigh distributed and, in particular, that each $\alpha_\ell\sim \mathcal{CN}(0,N_{\rm T}N_{\rm R}/\ell(D))$, where $\ell(D)$ denotes the average pathloss between TX and RX.  

In this work, we consider the temporally correlated mmWave channel model of \cite{HeTaHaKu14}, where the channel is assumed to be a block fading one, and the matrix $\mathbf{H}(n)$ at the $n$th channel block is derived from $\mathbf{H}(n-1)$ at the $(n-1)$th channel block ($\mathbf{H}$ for every block is obtained from \eqref{Eq:Channel_Model}) as $\mathbf{H}(n) = \mathbf{A}_{\rm R}\left(\boldsymbol{\theta}(n)\right){\rm diag}\{\boldsymbol{a}(n)\}\mathbf{A}_{\rm T}^{\rm H}\left(\boldsymbol{\phi}(n)\right)$. In the latter expression, $\boldsymbol{a}(n)$ denotes the gains' vector at the $n$th channel block that is modeled as
\begin{equation}\label{Eq:Evolution_Rayleigh}
\boldsymbol{a}(n) = \rho\boldsymbol{a}(n-1) + \sqrt{1-\rho^2}\boldsymbol{\beta}(n),
\end{equation}
where $\rho\in[0,1]$ represents the time correlation coefficient between the $n$th and $(n-1)$th channel blocks, and is given by
\begin{equation}\label{Eq:rho}
\rho=\frac{\mathbb{E}\{\alpha_\ell(n-1)\alpha_\ell^*(n)\}}{\sqrt{{\rm var}\{\alpha_\ell(n-1)\}{\rm var}\{\alpha_\ell^*(n)\}}}.
\end{equation}
According to the Jakes' model \cite{B:Jakes	}, $\rho=J_0(2\pi f_DT_{\rm bl})$ with $J_0(\cdot)$ denoting the zeroth order Bessel function of the first kind, $f_D$ being the maximum Doppler frequency, and $T_{\rm bl}$ is the block length of the channel. In \eqref{Eq:Evolution_Rayleigh}, $\boldsymbol{\beta}(n)\in\mathcal{C}^{L\times 1}$ is a diagonal matrix with each entries being independent of those in $\boldsymbol{a}(n-1)$ and drawn from  $\mathcal{CN}(0,N_{\rm T}N_{\rm R}/\ell(D))$. In addition, we assume that the matrices with the AoDs $\boldsymbol{\phi}(n)\triangleq[\phi_1(n)\,\phi_2(n)\,\cdots\,\phi_L(n)]$ and the AoAs $\boldsymbol{\theta}(n)\triangleq[\theta_1(n)\,\theta_2(n)\,\cdots\,\theta_L(n)]$ at the $n$th channel block $\mathbf{H}(n)$ can be modeled as
\begin{subequations}
\begin{equation}\label{Eq:Evolution_A_TX}
\mathbf{A}_{\rm T}\left(\boldsymbol{\phi}(n)\right) = \mathbf{A}_{\rm T}\left(\boldsymbol{\phi}(n-1) + \boldsymbol{\Delta\phi}\right),
\end{equation}
\begin{equation}\label{Eq:Evolution_A_RX}
\mathbf{A}_{\rm R}\left(\boldsymbol{\theta}(n)\right) = \mathbf{A}_{\rm R}\left(\boldsymbol{\theta}(n-1) + \boldsymbol{\Delta\theta}\right),
\end{equation}
\end{subequations}
where each entry of the $L$-dimension vectors $\boldsymbol{\Delta\phi}$ and $\boldsymbol{\Delta\theta}$ are uniformly distributed in $(-\delta,\delta)$ with $\delta$ being small. In general, in quite static environments for which mmWave communication seems to be a feasible communication paradigm, $\delta$ depends on the directional speed of TX and/or RX. 

\section{CS-Based mmWave Channel Estimation}\label{sec:CS_Channel_Estimation} 
In this section, similar to \cite{Alkhateeb_JSTSP}, we formulate the problem of channel estimation for the considered mmWave communication system as a CS problem. Suppose that TX uses a training beamforming vector $\mathbf{f}_p\in\mathcal{C}^{N_{\rm T}\times1}$ to transmit the unit-power training symbol $s_{\rm tr}\in\mathcal{C}$, and RX deploys a measurement vector $\mathbf{w}_q\in\mathcal{C}^{N_{\rm R}\times1}$ to process the received signal. Then, according to the system model described in \eqref{Eq:Estimated_Symbols}, the output signal at RX can be expressed as
\begin{equation}\label{Eq:Training_System_Model}
y_{q,p} = \mathbf{w}_q^{\rm H}\mathbf{H}\mathbf{f}_ps_{\rm tr} + \mathbf{w}_q^{\rm H}\mathbf{n}_{q,p},
\end{equation}
where $\mathbf{n}_{q,p}\in\mathcal{C}^{N_{\rm R}\times1}$ denotes the zero-mean AWGN vector at this training phase with covariance matrix $\sigma^2\mathbf{I}_{N_{\rm R}}$. 

If TX utilizes $M_{\rm T}$ training beamforming vectors $\mathbf{f}_p$ with $p=1,2,\ldots,M_{\rm T}$ at $M_{\rm T}$ successive time slots within each coherent channel block, and RX makes use of $M_{\rm R}$ measurement vectors $\mathbf{w}_q$ with $q=1,2,\ldots,M_{\rm R}$ at $M_{\rm R}$ successive time slots to process $s_{\rm tr}$ over each of the latter beamforming vectors, the following $M_{\rm R}\times M_{\rm T}$ complex-valued matrix may be constructed at RX
\begin{equation}\label{eq:matform}
\mathbf{Y} =\mathbf{W}^{\rm H}\mathbf{H}\mathbf{F}s_{\rm tr} + \mathbf{N},
\end{equation}
where $[\mathbf{Y}]_{q,p}\triangleq y_{q,p}$, $\mathbf{W}\triangleq[\mathbf{w}_1\,\mathbf{w}_2\,\cdots\,\mathbf{w}_{M_{\rm R}}]\in\mathcal{C}^{N_{\rm R}\times M_{\rm R}}$, $\mathbf{F}\triangleq[\mathbf{f}_1\,\mathbf{f}_2\,\cdots\,\mathbf{f}_{M_{\rm T}}]\in\mathcal{C}^{N_{\rm T}\times M_{\rm T}}$, and $\mathbf{N}\in\mathcal{C}^{M_{\rm R}\times M_{\rm T}}$ being a noise matrix, with each element defined as $[\mathbf{N}]_{q,p}\triangleq\mathbf{w}_q^{\rm H}\mathbf{n}_{q,p}$. By using the steps described in \cite[eqs. (12)-(15)]{Alkhateeb_JSTSP} and assuming that $s_{\rm tr}=\sqrt{P_{\rm tr}}$ with $P_{\rm tr}$ representing the average transmitted power in the training phase, \eqref{eq:matform} can be equivalently written in a vectorized fashion as follows
\begin{equation}\label{eq:vecform}
\mathbf{y}_{\rm v} \triangleq \sqrt{P_{\rm tr}}\left(\mathbf{F}^{\rm T}\!\otimes\!\mathbf{W}^{\rm H}\right)\left(\mathbf{A}_{\rm T}^*\left(\boldsymbol{\phi}\right)\circ\mathbf{A}_{\rm R}\left(\boldsymbol{\theta}\right)\right)\boldsymbol{\alpha} + \mathbf{n}_{\rm v},
\end{equation}
where $\mathbf{y}_{\rm v}\triangleq{\rm vec}(\mathbf{Y})\in\mathcal{C}^{M_{\rm T}M_{\rm R}\times1}$ and $\mathbf{n}_{\rm v}\triangleq{\rm vec}(\mathbf{N})\in\mathcal{C}^{M_{\rm T}M_{\rm R}\times1}$, with ${\rm vec}(\cdot)$ denoting the vectorization operation. The complex-valued matrix $\mathbf{A}_{\rm T}^*(\boldsymbol{\phi})\circ\mathbf{A}_{\rm R}(\boldsymbol{\theta})$ has the dimension $N_{\rm T}N_{\rm R}\times L$. Assuming that AoDs and AoAs are selected from large collections of angles (also termed as dictionaries) with $G_{\rm T}$ and $G_{\rm R}$ elements, respectively, such that $G_{\rm T},G_{\rm R}\gg L$, \eqref{eq:vecform} can be approximated as 
\begin{equation}\label{eq:vecform_1}
\mathbf{y}_{\rm v} \cong \sqrt{P_{\rm tr}}\left(\mathbf{F}^{\rm T}\mathbf{A}_{\rm T}^*\left(\bar{\boldsymbol{\phi}}\right)\otimes\mathbf{W}^{\rm H}\mathbf{A}_{\rm R}\left(\bar{\boldsymbol{\theta}}\right)\right)\mathbf{z} + \mathbf{n}_{\rm v}.
\end{equation}
In the latter expression, $\mathbf{A}_{\rm T}\left(\bar{\boldsymbol{\phi}}\right)\in\mathcal{C}^{N_{\rm T}\times G_{\rm T}}$, with $\bar{\boldsymbol{\phi}}\triangleq[\bar{\phi}_1\,\bar{\phi}_2\,\cdots\,\bar{\phi}_{G_{\rm T}}]$, and $\mathbf{A}_{\rm R}\left(\bar{\boldsymbol{\theta}}\right)\in\mathcal{C}^{N_{\rm R}\times G_{\rm R}}$, with $\bar{\boldsymbol{\theta}}\triangleq[\bar{\theta}_1\,\bar{\theta}_2\,\cdots\,\bar{\theta}_{G_{\rm R}}]$, are the dictionary matrices with the array response vectors for all the collections of angles. For collections taken from uniform grids, the elements of $\bar{\boldsymbol{\phi}}$ and $\bar{\boldsymbol{\theta}}$ are constructed as $\bar{\phi}_i=2\pi i/G_{\rm T}$ with $i=0,2,\ldots,G_{\rm T}-1$ and  $\bar{\theta}_j=2\pi j/G_{\rm R}$ with $j=0,2,\ldots,G_{\rm R} -1$, respectively. Finally, in \eqref{eq:vecform_1}, $\mathbf{z}\in\mathcal{C}^{G_{\rm T}G_{\rm R}\times1}$ includes the path gains for all combinations of quantized AoDs and AoAs. It is noted that, both $G_{\rm T}$ and $G_{\rm R}$ in \eqref{eq:vecform_1} are assumed to be sufficiently large such that \eqref{eq:vecform_1} tightly approximates \eqref{eq:vecform}.  

The mathematical representation in \eqref{eq:vecform_1} is a sparse formulation of the mmWave channel estimation problem due to the fact that $\mathbf{z}$ has only $L$ non-zero elements and by construction holds that $L\ll G_{\rm T}G_{\rm R}$. Hence, one may resort to the CS rationale for estimating the channel (i$.$e$.$, the AoD, AoA, and the gain of each of the $L$ propagation paths). The unknown channel can be, therefore, estimated by solving the following optimization problem  
\begin{align}\label{eq:CS}
\min_{\mathbf{z}}\left\|\mathbf{y}_{\rm v}-\bm{\Phi}\mathbf{z}\right\|^2\,\mathrm{s.t.}\,\left\|\mathbf{z}\right\|_0\leq L,
\end{align}
where $\bm{\Phi}\triangleq\sqrt{P_{\rm tr}}\left(\mathbf{F}^{\rm T}\mathbf{A}_{\rm T}^*\left(\bar{\boldsymbol{\phi}}\right)\otimes\mathbf{W}^{\rm H}\mathbf{A}_{\rm R}\left(\bar{\boldsymbol{\theta}}\right)\right)$ is a $M_{\rm T}M_{\rm R}\times G_{\rm T}G_{\rm R}$ complex-valued matrix. The intuition behind the optimization \eqref{eq:CS} is the following: we want to compute a solution that minimizes the error between the input and output vectors, and at the same time we restrict this solution to have at most $L$ non-zero entries (via the constraint including the $\ell_0$ norm). It is by now well established that the aforementioned optimization problem cannot be solved in polynomial time due to the $\ell_0$-norm constraint (see, e$.$g$.$, \cite{cande2008introduction} and references therein). To overstep this limitation, one could resort to the convex relaxation of the aforementioned problem, according to which the $\ell_0$ norm is substituted by the convex $\ell_1$ norm. An alternative option is to deploy greedy algorithms which, in essence, will search inside the available dictionary with the $G_{\rm T}G_{\rm R}$ combinations of quantized AoDs and AoAs for the actual $L$ propagation paths, and estimate their gains. A typical greedy algorithm for the estimation of the unknown sparse vector $\mathbf{z}$ in \eqref{eq:CS} is the compressive sampling matching pursuit (CoSaMP) algorithm \cite{needell2009cosamp}; its complexity will be discussed in the following section.

\section{Low Complexity CS Estimation\\ of Temporally Correlated mmWave Channels}\label{sec:lowOMP}
In this section, we aim at exploiting the characteristics of the considered temporally correlated mmWave channel model to design low complexity CS-based channel estimation techniques for mmWave systems comprising of TXs and RXs that employ antenna arrays with hybrid A/D processing capabilities. In particular, we present two algorithms that capitalize on the model described by \eqref{Eq:Evolution_A_TX} and \eqref{Eq:Evolution_A_RX} for the AoDs and AoAs, respectively, to reduce the complexity of CS-based mmWave channel estimation, without necessarily degrading its performance.

\subsection{Correlation-Aware CS Channel Estimation} 
Suppose that the AoDs $\boldsymbol{\phi}(n-1)$ and AoAs $\boldsymbol{\theta}(n-1)$ for all $L$ propagation paths of the $(n-1)$th channel block are perfectly estimated. According to the temporally correlated mmWave channel model presented in Section~\ref{Sec:Channel_Model}, at the $n$th channel block, it holds that $\boldsymbol{\phi}(n) = \boldsymbol{\phi}(n-1) + \boldsymbol{\Delta\phi}$ and $\boldsymbol{\theta}(n) = \boldsymbol{\theta}(n-1) + \boldsymbol{\Delta\theta}$. This indicates that, for each $\ell$th propagation path, $\phi_\ell(n)\in\mathcal{G}_{{\rm T},\ell}(n)$ with $\mathcal{G}_{{\rm T},\ell}(n)\triangleq[\phi_{\ell}(n-1)-\delta,\phi_{\ell}(n-1)+\delta]$ and $\theta_\ell(n)\in\mathcal{G}_{{\rm R},\ell}(n)$ with $\mathcal{G}_{{\rm R},\ell}(n)\triangleq[\theta_{\ell}(n-1)-\delta,\theta_{\ell}(n-1)+\delta]$. In other words, the angle sets $\mathcal{G}_{{\rm T},\ell}(n)$ and $\mathcal{G}_{{\rm R},\ell}(n)$ contain all the angles for the $\ell$th path of the $n$th channel block, the distance of which is at most $\delta$ from the estimated angles for this path at the $(n-1)$th channel block. By quantizing the angles within each of the sets $\mathcal{G}_{{\rm T},\ell}(n)$ and $\mathcal{G}_{{\rm R},\ell}(n)$, and assuming that $\bar{G}_{\rm T}(n)$ and $\bar{G}_{\rm R}(n)$ denote the cardinalities for each $\ell$ of $\mathcal{G}_{{\rm T},\ell}(n)$ and $\mathcal{G}_{{\rm R},\ell}(n)$, respectively, we can approximate $\mathbf{y}_{\rm v}$ in \eqref{eq:vecform_1} at the $n$th channel block as  
\begin{equation}\label{eq:vecform_1_n1}
\mathbf{y}_{\rm v}(n) = \hat{\bm{\Phi}}(n)\hat{\mathbf{z}}(n) + \mathbf{n}_{\rm v}(n),
\end{equation}
where $\hat{\bm{\Phi}}(n)\in\mathcal{C}^{M_{\rm T}M_{\rm R}\times L^2\bar{G}_{\rm T}(n)\bar{G}_{\rm R}(n)}$ and is given by
\begin{align} \label{eq:Phi_hat}
\hat{\bm{\Phi}}(n) &= \sqrt{P_{\rm tr}}\left(\mathbf{F}^{\rm T}\mathbf{A}_{\rm T}^*(\hat{\boldsymbol{\phi}}(n))\otimes\mathbf{W}^{\rm H}\mathbf{A}_{\rm R}(\hat{\boldsymbol{\theta}}(n))\right) \\
\nonumber & = \sqrt{P_{\rm tr}}\left(\mathbf{F}^{\rm T}\!\otimes\!\mathbf{W}^{\rm H}\right) 
\left(\mathbf{A}_{\rm T}^*(\hat{\boldsymbol{\phi}}(n))\circ\mathbf{A}_{\rm R}(\hat{\boldsymbol{\theta}}(n))\right).
\end{align}
In \eqref{eq:Phi_hat}, the vector $\hat{\boldsymbol{\phi}}(n)$ includes the quantized AoDs and has dimension $L\bar{G}_{\rm T}(n)$, whereas the vector $\hat{\boldsymbol{\theta}}(n)$ contains the quantized AoAs and is of dimension $L\bar{G}_{\rm R}(n)$. Moreover, $\mathbf{A}_{\rm T}(\hat{\boldsymbol{\phi}}(n))$ has the dimension $N_{\rm T}\times L\bar{G}_{\rm T}(n)$, $\mathbf{A}_{\rm R}(\hat{\boldsymbol{\theta}}(n))$ is of dimension $N_{\rm R}\times L\bar{G}_{\rm R}(n)$, and the column vector $\hat{\mathbf{z}}$ with dimension $L^2\bar{G}_{\rm T}(n)\bar{G}_{\rm R}(n)$ includes the path gains for the all the combinations of quantized AoDs and AoAs. Note that, in the last form of \eqref{eq:Phi_hat}, the matrix $\mathbf{F}^{\rm T}\!\otimes\!\mathbf{W}^{\rm H}$ with dimension $M_{\rm T}M_{\rm R}\times N_{\rm T}N_{\rm R}$ does not depend on $n$, and hence, needs to be computed only once. 
\begin{algorithm}[!t]
\caption{Correlation-Aware CS Channel Estimation}\label{Reduce_CoSaMP}
\begin{algorithmic}[1]
\STATE \textbf{Initialization:} Solve \eqref{eq:CS} with the $M_{\rm T}M_{\rm R}\times G_{\rm T}G_{\rm R}$ matrix $\bm{\Phi}(1)$ using CoSaMP to compute the initial $\mathbf{z}(1)$ for the $1$st channel block.\\
\textbf{FOR $n=2,3,\ldots$}
\STATE Construct the sets $\mathcal{G}_{{\rm T},\ell}(n)$ and $\mathcal{G}_{{\rm R},\ell}(n)$ for each $\ell$th path using $\delta$ and the estimated AoDs and AoAs from the $(n-1)$th channel block.
\STATE Form the vectors $\hat{\boldsymbol{\phi}}(n)$ and $\hat{\boldsymbol{\theta}}(n)$ with the quantized AoDs and AoAs, respectively. 
\STATE Compute $\mathbf{A}_{\rm T}(\hat{\boldsymbol{\phi}}(n))$ and $\mathbf{A}_{\rm R}(\hat{\boldsymbol{\theta}}(n))$, and calculate the $M_{\rm T}M_{\rm R}\times L^2\bar{G}_{\rm T}(n)\bar{G}_{\rm R}(n)$ matrix $\hat{\bm{\Phi}}(n)$ using \eqref{eq:Phi_hat}.\\
\STATE Substitute \eqref{eq:vecform_1_n1} and \eqref{eq:Phi_hat} into \eqref{eq:CS} and solve using CoSaMP to compute $\hat{\mathbf{z}}(n)$ for the $n$th channel block.\\
\textbf{ENDFOR}
\end{algorithmic}
\end{algorithm}

By plugging \eqref{eq:vecform_1_n1} and \eqref{eq:Phi_hat} into \eqref{eq:CS} for the estimation of the $n$th channel block, it becomes apparent that, compared with the substitution of \eqref{eq:vecform_1} into \eqref{eq:CS} for the same channel block, the dimensionality of the optimization problem can be significantly reduced if $L^2\bar{G}_{\rm T}(n)\bar{G}_{\rm R}(n)\ll G_{\rm T}G_{\rm R}$. This condition is highly probable to hold since, in general, $L$ is usually very small in mmWave channels; the same happens for both $\bar{G}_{\rm T}(n)$ and $\bar{G}_{\rm R}(n)$ $\forall$$n$ due to the fact that $\delta$ is assumed to be small. On the contrary, $G_{\rm T}G_{\rm R}$ represents the size of the available angles' dictionary which, in principal, needs to be very large. The first proposed low complexity CS-based channel estimation algorithm that exploits the temporal correlation of the mmWave channel is summarized in Algorithm~\ref{Reduce_CoSaMP}. Initially, the $1$st channel block is estimated by using the CoSaMP algorithm to solve \eqref{eq:CS} after replacing \eqref{eq:vecform_1}, i$.$e$.$, using the whole available dictionary of angles. Then, for each $n$th channel block, the sets $\mathcal{G}_{{\rm T},\ell}(n)$ and $\mathcal{G}_{{\rm R},\ell}(n)$ with the feasible AoDs and AoAs, respectively, for each $\ell$th path are formulated. In doing so, each $\ell$th path's AoD and AoA estimation from the $(n-1)$th channel block is used together with the $\delta$ value. Note that, in practice, $\delta$ must be estimated. For example, in a point-to-point mmWave communication system with a strong line-of-sight component where only the RX moves, $\delta$ can be obtained from RX's directional velocity or new position at the $n$th channel block \cite{Taranto_SPmag2014}. By quantizing the sets $\mathcal{G}_{{\rm T},\ell}(n)$ and $\mathcal{G}_{{\rm R},\ell}(n)$ for each $\ell$, $\hat{\boldsymbol{\phi}}(n)$ and $\hat{\boldsymbol{\theta}}(n)$ are obtained, which are then used to compute $\mathbf{A}_{\rm T}(\hat{\boldsymbol{\phi}}(n))$ and $\mathbf{A}_{\rm R}(\hat{\boldsymbol{\theta}}(n))$. The latter matrices are substituted into \eqref{eq:Phi_hat} and then in \eqref{eq:vecform_1_n1}, and finally, the latter expression is plugged into \eqref{eq:CS} to obtain $\hat{\mathbf{z}}(n)$ using the greedy algorithm CoSaMP.

\subsection{Sparsity-Aware Iterative CS Channel Estimation}
As previously described, Algorithm~\ref{Reduce_CoSaMP} makes use of the CoSaMP algorithm to estimate the mmWave channel at every channel block. For the initial channel block estimation, the unknown vector is of size equal to the size of the whole dictionary of angles, while, for the subsequent channel blocks, the size of the unknown vector is significantly reduced by exploiting the time correlation between blocks. However, in Algorithm~\ref{Reduce_CoSaMP}, the channel estimation at each block does not take into account the estimation of previous blocks. We, therefore, present in the following an iterative algorithm for mmWave channel estimation that capitalizes on the sparsity of the unknown vector and computes it for each channel block using knowledge of the estimation of previous blocks. More specifically, at each channel block, the computed estimate of the unknown vector at the previous block is used as a warm start. This algorithm has the following two notable benefits: \textit{i}) no matrix inversions are required when computing $\hat{\mathbf{z}}(n)$ for the $n$th channel block; and \textit{ii}) one can trade-off complexity and accuracy by adjusting the number of iterations for calculating $\hat{\mathbf{z}}(n)$ for the $n$th channel block.
\begin{algorithm}[!t]
\caption{Sparsity-Aware Iterative CS Channel Estimation}\label{Sparcity_Aware}
\begin{algorithmic}[1]
\STATE \textbf{Initialization:} Compute the initial $\mathbf{z}(1)$ using the Step $1$ of Algorithm~\ref{Reduce_CoSaMP}.\\
\textbf{FOR $n=2,3,\ldots$}
\STATE Construct the sets $\mathcal{G}_{{\rm T},\ell}(n)$ and $\mathcal{G}_{{\rm R},\ell}(n)$ for each $\ell$th path as in the Step $2$ of Algorithm~\ref{Reduce_CoSaMP}.
\STATE Form the vectors $\hat{\boldsymbol{\phi}}(n)$ and $\hat{\boldsymbol{\theta}}(n)$ as in the Step $3$ of Algorithm~\ref{Reduce_CoSaMP}. 
\STATE Calculate the matrix $\hat{\bm{\Phi}}(n)$ using the Step $4$ of Algorithm~\ref{Reduce_CoSaMP}.\\
\textbf{FOR $k=1,2,\ldots, I$}
\STATE Compute $\hat{\mathbf{z}}_k(n)$ using \eqref{eq:z_iterations}.\\
\textbf{ENDFOR} \\
\STATE Set $\hat{\mathbf{z}}(n)=\hat{\mathbf{z}}_I(n)$ and $\hat{\mathbf{z}}_0(n+1)=\hat{\mathbf{z}}(n)$.\\
\textbf{ENDFOR}
\end{algorithmic}
\end{algorithm}

The sparsity-aware iterative channel estimation algorithm is summarized in Algorithm~\ref{Sparcity_Aware}. The first four steps of this algorithm are the same with those of Algorithm~\ref{Reduce_CoSaMP}. Instead of using CoSaMP to estimate $\hat{\mathbf{z}}(n)$ for $n\geq2$, Algorithm~\ref{Sparcity_Aware} deploys the so-called iterative hard thresholding (IHT) algorithm, originally proposed in \cite{blumensath2009iterative}. It is noted that other iterative algorithms could be used as well, such as the soft thresholding algorithm \cite{bredies2008linear}. With IHT, the unknown vector at the $n$th channel block is estimated at the $k$th iteration ($k=1,2,\dots,I$ with $I$ denoting the total number of iteration) as 
\begin{equation}\label{eq:z_iterations}
\hat{\mathbf{z}}_k(n) = \mathbb{H}_L\left\{\hat{\mathbf{z}}_{k-1}(n) + \hat{\bm{\Phi}}^*(n)\left(\mathbf{y}_{\rm v}(n) - \hat{\bm{\Phi}}(n)\hat{\mathbf{z}}_{k-1}(n)\right)\right\},
\end{equation}
where $\mathbb{H}_L\left\{\cdot\right\}$ is a non-linear operator that keeps the $L$ largest in amplitude components of a vector and sets the remaining ones to zero. Finally, the estimation of the $n$th channel block is set to $\hat{\mathbf{z}}_I(n)$. For the estimation of the $(n+1)$th channel block, we initialize as $\hat{\mathbf{z}}_0(n+1)=\hat{\mathbf{z}}(n)$.

\subsection{Complexity Comparison}\label{Sec:Complexities}
With the Algorithms~\ref{Reduce_CoSaMP} and~\ref{Sparcity_Aware} we aim at exploiting the highly probable temporal correlation of mmWave channels in order to reduce the complexity of their estimation, while, in parallel, certifying an acceptable level of performance. By assuming that we estimate $B$ successive channel blocks, and that $\bar{G}_{\rm T}(n)=\bar{G}_{\rm T}$ and $\bar{G}_{\rm R}(n)=\bar{G}_{\rm R}$ $\forall$$n$, we next list the computational complexities of the two proposed algorithms and that of a fully greedy algorithm that uses the CoSaMP algorithm for the whole angles' dictionary at the estimation of each channel block.    
\begin{itemize}
\item Full Greedy Algorithm: With this algorithm, the CoSaMP is deployed $B$ times and the complexity equals to $O\left(BK(G_{\rm T}G_{\rm R}(M_{\rm T}M_{\rm R}+1)+2M_{\rm T}M_{\rm R})\right)$, where $K$ is the number of CoSaMP iterations.
\item Algorithm~\ref{Reduce_CoSaMP}: With this algorithm, the complexity equals to
$O\left(BK(L^2\bar{G}_{\rm T}\bar{G}_{\rm R}(M_{\rm T}M_{\rm R}+1)+2M_{\rm T}M_{\rm R})\right)$. 
\item Algorithm~\ref{Sparcity_Aware}: The complexity in this case further reduces to $O\left(BIL^2\bar{G}_{\rm T}\bar{G}_{\rm R}(M_{\rm T}M_{\rm R}+1)\right)$. 
\end{itemize}

\section{Numerical Results and Discussion}\label{sec:Results}
The performance of the proposed low complexity channel estimation algorithms suitable for temporally correlated mmWave channels is investigated in this section in terms of: \textit{i}) the MSE of channel estimation; and \textit{ii}) the mean achievable rate of a point-to-point mmWave system with large linear hybrid A/D antenna arrays at both communication ends. Without loss of generality, we will present results for strong line-of-sight mmWave channels having only one propagation path, i$.$e$.$, $L=1$, leaving investigations for $L>1$ propagation paths for the extended version of this paper.    

\subsection{MSE of Channel Estimation}
\begin{figure}[!t]
\centering
\includegraphics[width=0.42\textwidth]{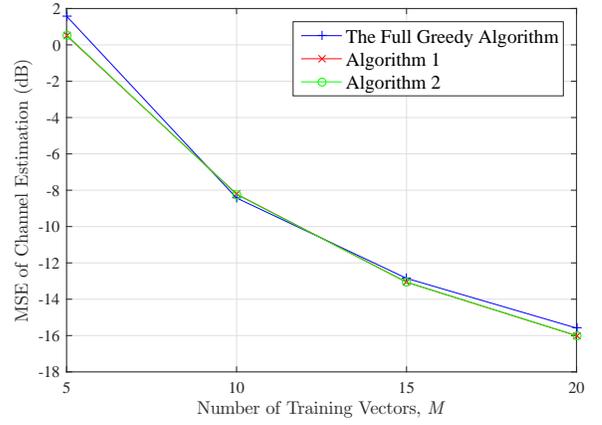}
\caption{The MSE of mmWave channel estimation in dB as a function of the number of training vectors $M$ for the high noise scenario.}
\label{Fig:MSE_High_Noise}
\end{figure}

\begin{figure}[!t]
\centering
\includegraphics[width=0.42\textwidth]{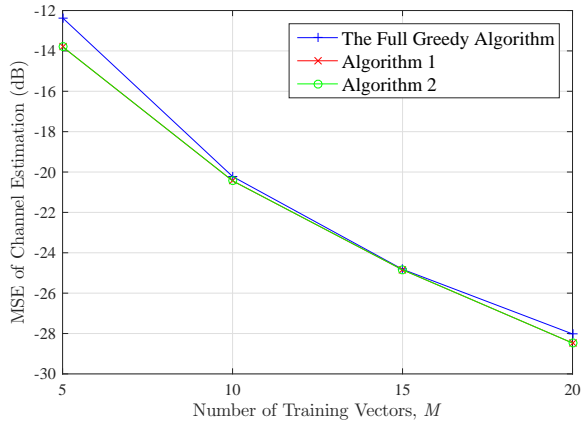}
\caption{The MSE of mmWave channel estimation in dB as a function of the number of training vectors $M$ for the low noise scenario.}
\label{Fig:MSE_Low_Noise}
\end{figure}

We compare the MSE performance of channel estimation with the two proposed low complexity algorithms with that of the full greedy algorithm for two different noise scenarios: \textit{i}) a high noise scenario, where the SNR in the estimation phase is $-10$ dB; and \textit{ii}) a low noise scenario, with the SNR being $0$ dB. For TX we assume that $N_{\rm T}=32$ and $N_{\rm R}=64$ is considered for RX. Channel estimation is performed over $20$ consecutive coherent channel blocks. Parameter $\delta$ is set to $3^{\rm o}$ and $\rho$ is set to $0.8$. We assume that $M_{\rm T}=M_{\rm R}=M$ for the training beamforming and measurement vectors. We also consider that, at each channel block, the dimension of the $\bm{\Phi}(n)$ matrix $\forall$$n=1,2,\ldots,20$ for the full greedy algorithm is $M^2\times2048$, whereas for both the proposed low complexity algorithms $\hat{\bm{\Phi}}(n)$ has the dimension $M^2\times20$. Finally, for Algorithm~\ref{Sparcity_Aware}, we have set $I=10$. In Figs$.$~\ref{Fig:MSE_High_Noise} and~\ref{Fig:MSE_Low_Noise}, we vary the number of training vectors $M$ and compute the MSE performance of channel estimation as 
\begin{equation}\label{eq:MSE}
{\rm MSE} = 20^{-1}\sum_{n=1}^{20} \|\mathbf{z}(n)- \tilde{\mathbf{z}}(n)\|^2,
\end{equation}
where $\mathbf{z}(n)$ represents the true sparse vector and $\tilde{\mathbf{z}}(n)$ denotes its estimate. For the full greedy algorithm, $\mathbf{z}(n)$ at each $n$th channel block is the one appearing in \eqref{eq:vecform_1}, whereas for the two proposed low complexity algorithms, $\mathbf{z}(n)$ at each $n$th channel block coincides with $\hat{\mathbf{z}}(n)$, i$.$e$.$, the one appearing in \eqref{eq:vecform_1_n1}. Since we have considered $L=1$, $\mathbf{z}(n)$ and $\tilde{\mathbf{z}}(n)$ for all algorithms will have exactly one non-zero element, which needs to be as close as possible to the non-zero element of $\boldsymbol{a}$. In addition, from the position of this element in $\tilde{\mathbf{z}}(n)$, the pair of AoD and AoA for the sole propagation path will be extracted, thus, providing the entire channel state information. As seen from both Figs$.$~\ref{Fig:MSE_High_Noise} and~\ref{Fig:MSE_Low_Noise}, the proposed algorithms not only reduce significantly the complexity of channel estimation, but also exhibit a slightly improved performance than that of the full greedy algorithm. This behavior is a direct consequence of the fact that we reduce our solution space, since the a priori knowledge coming from our model limits the focus on a smaller set of angles into which the solution lies. Furthermore, as expected, the more measurements we utilize the better the channel estimation accuracy is. Last but not least, the performance of the two proposed algorithms is similar, which is expected since the iterative Algorithm~\ref{Sparcity_Aware} converges to the solution of Algorithm~\ref{Reduce_CoSaMP} for a sufficient number of iterations. 

\subsection{Mean Achievable Rate}
\begin{figure}[!t]
\centering
\includegraphics[width=0.42\textwidth]{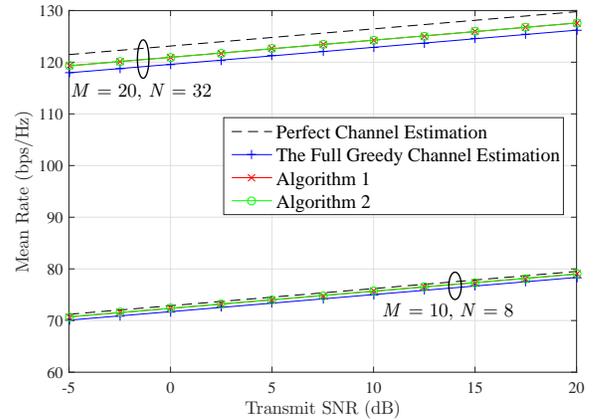}
\caption{The mean achievable rate versus the transmit SNR in dB with CS-based channel estimation and beamforming for different numbers of training vectors $M$ and antennas $N$.}
\label{Fig:Ergodic_Rate}
\end{figure}

Consider a TX communicating with a mobile RX, where both are incapable of performing BB processing and are equipped with single-RF (i$.$e$.$, $N_{\rm RF}=1$) $N$-element ULAs of half-wavelength inter-antenna spacing. RX is assumed to move so that $\delta=3^{\rm o}$ between consecutive channel blocks in the model described by \eqref{Eq:Evolution_A_TX} and \eqref{Eq:Evolution_A_RX}, and $\rho=0.9037$ \cite{HeTaHaKu14}. Suppose also that, RX estimates the matrix $\mathbf{H}(n)$ at each $n$th coherent channel block as described in Sec$.$~\ref{sec:CS_Channel_Estimation}, using any of the algorithms presented in Sec$.$~\ref{sec:lowOMP}, and then sends this information through an ideal feedback channel to TX. Then, to establish communication, TX computes $\mathbf{V}(n)\triangleq\mathbf{V}_{\rm RF}(n)\in\mathcal{C}^{N\times 1}$ and RX calculates $\mathbf{U}(n)\triangleq\mathbf{U}_{\rm RF}(n)\in\mathcal{C}^{N\times 1}$ as the right- and left-singular vectors of the estimate for $\mathbf{H}(n)$, respectively. In Fig$.$~\ref{Fig:Ergodic_Rate}, we depict the mean achievable rate over $100$ channel blocks for all the considered channel estimation algorithms ($G_{\rm T}=G_{\rm R}=10^3$ and $\bar{G}_{\rm T}(n)=\bar{G}_{\rm R}(n)=10$ $\forall$$n$) as a function of the transmit SNR. Within this figure, we vary the values of the parameters $M$ and $N$, and as shown, both the proposed low complexity algorithms result in mean rate that is sufficiently close to that of perfect channel estimation. As expected, when increasing the number of antennas $N$, the number of required training vectors $M$ (i$.$e$.$, the number of measurements) needs to increase in order to achieve mean rate performance close to the perfect estimation case.     


\section{Conclusion}\label{sec:Conclusion}
In this paper, we focused on low complexity CS-based mmWave channel estimation, and presented a greedy and an iterative algorithms that capitalize on the potential temporal correlation of mmWave channels. We highlighted the impact of some key parameters on the accuracy of channel estimation.

\bibliographystyle{IEEEtran}
\bibliography{IEEEabrv,refs}

\end{document}